# DNA topology dictates emergent bulk elasticity and hindered macromolecular diffusion in DNA-dextran composites

Pawan Khanal, Karthik R Peddireddy, Juexin Marfai, Ryan McGorty, Rae M Robertson-Anderson*

*Department of Physics and Biophysics, University of San Diego, 5998 Alcala Park, San Diego, CA 92110*

*randerson@sandiego.edu

## Abstract

Polymer architecture plays critical roles in both bulk rheological properties and microscale macromolecular dynamics in entangled polymer solutions and composites. Ring polymers, in particular, have been the topic of much debate due to the inability of the celebrated reptation model to capture their observed dynamics. Macrorheology and differential dynamic microscopy (DDM) are powerful methods to determine entangled polymer dynamics across scales, yet they typically require different samples under different conditions, preventing direct coupling of bulk rheological properties to the underlying macromolecular dynamics. Here, we perform macrorheology on composites of highly-overlapping DNA and dextran polymers, focusing on the role of DNA topology (rings versus linear chains) as well as the relative volume fractions of DNA and dextran. On the same samples under the same conditions, we perform DDM and single-molecule tracking on embedded fluorescent-labeled DNA molecules immediately before and after bulk measurements. We show DNA-dextran composites exhibit unexpected non-monotonic dependences of bulk viscoelasticity and molecular-level transport properties on the fraction of DNA comprising the composites, with characteristics that are strongly dependent on the DNA topology. We rationalize our results as arising from stretching and bundling of linear DNA versus compaction, swelling, and threading of rings driven by dextran-mediated depletion interactions.

## 1. Introduction

Ring polymers have been the topic of fervent investigation for decades due to their intriguing rheological and dynamical properties, biological significance, and industrial applications. For example, DNA naturally occurs in ring formation, and conversion between supercoiled and open circular (ring) topology plays a critical role in DNA replication and repair [1-3]. Further, ring polymers can tune the rheological properties of polymeric blends for commercial and industrial use [4-6]. While the dynamics of entangled linear polymers are well described by the reptation model developed by de Gennes and Doi and Edwards [7, 8], its extension to ring polymers is not straightforward due to their lack of free ends [9-11]. The extent to which ring polymers form entanglements and corresponding entanglement plateaus, the effect and persistence of threading of one polymer by another, and the relaxation modes available to ring polymers remain topics of debate [12-17].

While previous rheological studies have shown that entangled ring polymers do not display entanglement plateaus that their linear counterparts do, [12, 13, 18-20] indicating weaker entanglements, other studies have shown that rings undergo very slow relaxation compared to entangled linear chains, such that the final scaling of the storage and loss moduli in the terminal flow regime, $G'(\omega) \sim \omega^2$ and $G''(\omega) \sim \omega^1$, is not reached [12, 17, 21-24]. Entangled rings are predicted to adopt double-folded or amoeba-like branched structures that undergo modified reptation that is faster than their linear counterparts [9, 25-28]. At the same time, the reported slow relaxation has been attributed to ring-ring threading, which has been predicted to lead to glassy dynamics from kinetically arrested states that exhibit heterogeneous transport [29, 30].

Because nearly all synthesis techniques used to produce solutions and melts of ring polymers result in a small percentage of linear polymer 'contaminants', threading of rings by linear contaminants likely also contribute to the reported slow relaxation of nominally 'pure' entangled rings [31]. The dominant relaxation mode of rings threaded by linear chains is constraint release, whereby a polymer relaxes stress by the threaded polymer unthreading itself and releasing its constraint [14, 22, 31], rather than the faster mode of reptation that otherwise dominates entangled polymer systems [8, 27].

The different rheological properties of entangled rings and linear polymers also manifests in the frequency dependence of the shear viscosity $\eta(\omega) \sim \omega^{-\gamma}$, an indicator of entanglements and how well polymers are able to align with shear flow. Highly entangled flexible linear polymer melts exhibit shear thinning $\eta(\omega) \sim \omega^{-\gamma}$ with $\gamma \approx 0.9\text{-}1$ [8, 20], whereas entangled linear DNA solutions in good solvent conditions exhibit slightly reduced thinning with $\gamma \approx 0.6 - 0.9$ [16, 32, 33]. Entangled ring melts and solutions have been reported to exhibit weaker shear thinning than their linear counterparts ($\gamma \approx 0.4 - 0.6$) owing to their reduced ability to conformationally align with shear flow [16, 20].

In ring-linear blends, threading of rings by linear chains have been shown to have profound effects on the rheology and dynamics [12, 21, 22, 31, 34-36], including emergent entanglement plateaus, increased zero-shear viscosity and shear-thinning, suppressed relaxation, and hindered diffusion as compared to monodisperse systems of linear chains or rings [16, 36-40]. For example, previous studies have shown that a small fraction of linear chains (>0.05) can cause entangled ring melts to transition from power-law stress relaxation [21, 41] to exhibiting entanglement plateaus comparable to their linear counterparts [19]. At the same time, a small fraction of rings (up to ~0.3) added to entangled linear polymer melts has been shown to increase the viscosity up to ~2-fold [14, 22]. Further, recent simulations have shown that rings in a melt of very short linear polymers (up to ~10-fold shorter than the rings) exhibit more swollen conformations, due to excess chain-end free volume effects, and diffuse faster than in their own melt [40]. However, when the length of the surrounding linear chains exceeds the entanglement length, ring diffusion drops substantially and swelling is reduced. Finally, the addition of linear chains to entangled rings increases the shear thinning exponent to be comparable to or larger than that for entangled linear polymers with reported values of $\gamma \approx 0.7 - 0.9$ [16, 19], suggesting that threading aids in stretching and aligning rings along the flow direction.

While the role of polymer topology in entangled melts, solutions and blends continues to be widely investigated, how topology modulates the rheology and dynamics of composites of two distinct polymeric species with different sizes, stiffnesses and structures remains scarcely explored [42]. Free energy minimization of composites, typically comprising a polymer matrix and a 'filler' species, leads to emergent behavior, such as non-monotonic dependence of viscoelastic and transport properties on the relative fraction of each species [43-47]. Emergent phenomena in composites can arise from entropically-driven polymer bundling and flocculation [42, 46, 48], enhanced miscibility and solubility [42, 47, 49, 50], or even liquid-liquid phase separation [51, 52].

For example, we previously showed that composites of flexible DNA and ~200-fold stiffer actin filaments exhibit enhanced stiffening and suppressed relaxation compared to networks of DNA or actin alone due to small-scale actin bundling, driven by DNA, that stiffens and strengthens the composite [46]. This phenomenon arises from depletion interactions, ubiquitous in biology, in which macromolecules of one species (actin) are driven together to maximize the available volume (and thus entropy) of the other smaller, faster and/or more abundant species (DNA). The unexpected non-monotonic dependence of composite stiffening on the actin fraction is a result of competition between actin bundling, which stiffens the actin network and suppresses thermal fluctuations, and actin network connectivity required to provide a percolated scaffold to support and entangle with the DNA. Similar non-monotonic emergence has been reported in crosslinked composites of cytoskeleton filaments [43, 47, 53].

Another recent study investigating the role of short chain additives to composites of long polymers and nanoparticles (NPs) reported a non-monotonic dependence of composite strength on the ratio of short and long polymer chains in polymer nanocomposites [54]. The peak in mechanical strength, which occurs at ~40% short chains, is attributed to an interplay between short chain adsorption on NPs–which induces stretching and swelling of the long chains and an associated increase in their overlap density–and reduced polymer entanglements and NP connectivity as the number of long NP-bridging polymers is reduced. The interplay between enhanced polymer self-association and network connectivity is similar in spirit to that reported in other composite systems [42, 46, 55].

Similar to Ref [54], we previously found stretching of tracer linear DNA from random coils to elongated snake-like conformations, when crowded by small dextran polymers at sufficient dextran volume fraction (~30%, $11c^*$) [56]. Conversely, ring DNA condensed into more compact spherical conformations under the same conditions. Both elongation and compaction are volume-minimizing conformations, driven by entropy maximization of the dextran crowders (i.e., the depletion interaction).

The complex and intriguing effects of polymer end-closure on rheology and macromolecular dynamics [10, 18, 20, 57], combined with the rich emergent phenomena that polymer composites exhibit [44, 55, 58-60], motivated us to examine the molecular-level transport and bulk rheological properties of composites of DNA and dextran–focusing on the role of DNA end-closure (Fig 1).

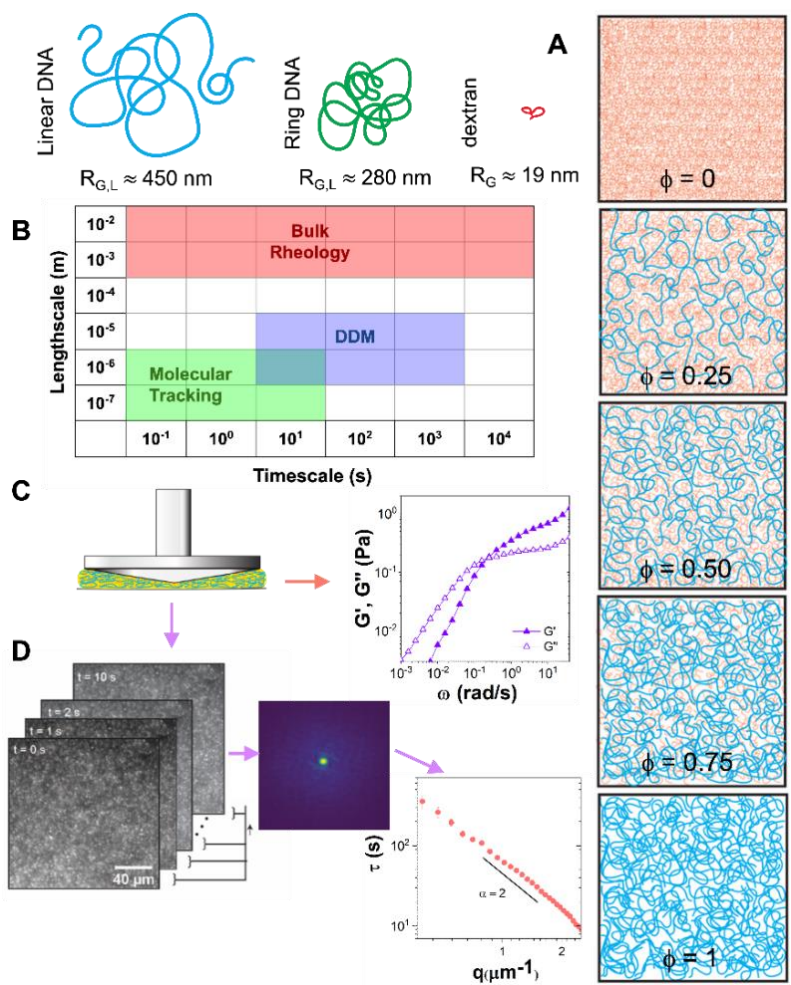

**Fig 1. Experimental platform to investigate bulk rheology and molecular-level dynamics in composites of DNA and dextran.** (**A**) Cartoon of macromolecules that comprise the composites, with their corresponding radius of gyration $R_G$ listed: 25-kbp linear DNA (blue), 25-kbp ring DNA (green) and 500 kDa dextran (red). The panel on the right depicts $11c^*$ composites of linear DNA and dextran at all DNA volume fractions $\phi$ that we investigate: $\phi = 0$ (100% dextran), $\phi = 0.25$ (25% DNA, 75% dextran), $\phi = 0.50$ (50% DNA, 50% dextran), $\phi = 0.75$ (75% DNA, 25% dextran), and $\phi = 1$ (100% DNA). (**B**) Lengthscales and timescales probed by the different techniques we use: bulk rheology, differential dynamic microscopy (DDM), and molecular-tracking. (**C**) We use a DHR3 rheometer to measure bulk viscoelastic properties of composites, including the elastic and viscous moduli $G'(\omega)$ (filled symbols) and $G''(\omega)$ (open symbols). (**D**) Immediately before and after bulk rheology measurements, we capture time-series of labeled DNA molecules diffusing in the sample using the fluorescence microscopy capabilities of the rheometer. We perform DDM analysis on captured time-series to determine the image structure function $D(q, \Delta t)$ as a function of lag time, $\Delta t$, and wavevector, $q$. By fitting $D(q, \Delta t)$, we determine the density fluctuation decay times $\tau$ as a function of $q$ to describe the DNA dynamics. Data shown in (C) and (D) is for linear DNA at $\phi = 1$.

Our work elucidates the effect of DNA topology (ring versus linear) and relative fraction of DNA and dextran on the dynamics of DNA-dextran composites from the scale of single polymers (~$10^2$ nm) to that of the macroscopic bulk (~cm). Specifically, we couple macrorheology with fluorescence microscopy, differential dynamic microscopy (DDM) and single-molecule tracking to directly connect the bulk rheological properties of the composites to the microscale transport properties of the comprising DNA.

While both macro- and micro-rheological techniques have been extensively used to investigate entangled polymers and other soft materials, these distinct measurements are typically performed on different samples with different preparation methods, chamber geometries, and sample volumes [61, 62]. As such, direct connection between the properties at these two scales is non-trivial [33, 63-65]. We overcome these limitations by performing imaging and rheology measurements in the same sample using a rheometer with high-speed fluorescence imaging capabilities. Further, we track DNA molecules comprising the composites, rather than embedded probes (as is typically done in microrheology experiments [66]), to directly report macromolecular dynamics via DDM.

Our results reveal a surprising non-monotonic dependence of the rheological properties on the fraction of DNA comprising the composites, with 75%-25% DNA-dextran composites exhibiting uniquely suppressed dissipation coupled with prolonged relaxation times and subdiffusion. We argue that this emergent behavior, that is strongly dependent on DNA topology, arises from entropically-driven self-association and stretching of linear DNA versus swelling and threading of ring DNA.

## 2. Materials and Methods

**2.1 DNA:** We prepare solutions of double-stranded DNA, 25 kbp in length, *via* replication of fosmid constructs in *Escherichia coli* followed by extraction, purification and concentrating as described previously [67-69]. Briefly, to replicate DNA, *E. coli* cultures containing the fosmid clone is grown from frozen glycerol stocks. To extract the DNA, cells are lysed via treatment with an alkaline solution. The extracted DNA is then renatured via treatment with an acidic detergent, precipitated in isopropanol, washed with 70% ethanol, and resuspended in TE10 buffer (10 mM Tris-HCl (pH 8), 1 mM EDTA, 10 mM NaCl). To purify the DNA, the solution is treated with Rnase A (to remove contaminating RNA) followed by phenol-chloroform extraction and dialysis (to remove proteins). We assess purity using UV absorbance and gel electrophoresis [68]. The purified DNA solution has a 260/280 absorbance ratio of ~1.8, in the accepted range for pure DNA with minimal protein contaminants [70], and no detectable RNA (which manifests as a low MW smear on a gel) (Fig S1).

We use gel electrophoresis to determine a concentration of 3.25 mg/mL of our purified solution, consisting of ~90% relaxed circular (ring) and ~10% linear DNA (Fig S1). Gel analysis is performed using Life Technologies E-Gel Imager and Gel Quant Express software (Fig S1). We convert half of the purified stock DNA solution to linear topology via treatment with the restriction enzyme ApaI (New England Biolabs). We confirm complete conversion to linear topology via gel electrophoresis (Fig S1).

The radii of gyration for the ring and linear DNA are $R_{G,R} \simeq 280$ nm and $R_{G,L} \simeq 450$ nm, respectively [69]. We compute polymer overlap concentrations for the linear and 90% ring DNA solutions via $c^* = (3/4\pi)(M/N_A)/(f_L R_{G,L}^3 + (1-f_L)R_{G,R}^3)$ where $M$ the molecular weight and $f_L$ is the fraction of linear DNA in the sample ($f_L = 1$ and 0.1 for linear and ring solutions, respectively) [8, 71], resulting in $c^*_{0.9R} \simeq 220$ µg/mL and $c^*_L \simeq 71$ µg/mL for ring and linear DNA, respectively. To image DNA for DDM and particle-tracking, we fluorescent-label a fraction of the DNA molecules with MFP488 (Mirus) using the manufacturer-supplied *Label* IT Labeling Kit and corresponding protocols (Mirus). The excitation/emission spectrum for MFP488 is 501/523 nm and the dye molecule to DNA base pair ratio is 5:1.

**2.2 Dextran:** We use molecular biology grade dextran (Fisher BioReagents BP1580100, Lot #196289), with molecular weight 500 kDa and $R_G \simeq 19$ nm [72]. The manufacturer-provided certificate of analysis shows suitability for protein fractionation and gel permeation chromatography, with heavy metal impurities and ignition residues below industry standards for molecular biology grade purity. To prepare composites, dextran is dissolved in TE10 buffer at a concentration of 28.9 mg/ml ($11c^*$) and homogenized via slow rotation at room temperature for >24 hrs. In the present study, as well as our previous studies investigating the diffusion of tracer DNA in crowded dextran solutions [56], we see no signs of adsorption or other non-steric interactions between dextran and DNA.

**2.3 Sample Preparation:** We prepare each sample at a volume of 350 µL comprising different volume fractions of DNA and dextran solutions each at $11c^*$, corresponding to 2.42, 0.78 and 28.9 mg/ml for ring DNA, linear DNA and dextran, respectively. We add 2 µL of labeled ring or linear DNA tracers to each sample for microscopy measurements. Each sample is prepared at least 4 days prior to experiments and rotated at 4ºC to mix and equilibrate. An oxygen scavenging system (45 µg/mL glucose, 43 µg/mL glucose oxidase, 7 µg/mL catalase, 5 µg/mL β-mercaptoethanol) is added to inhibit photobleaching. Composites comprising both DNA and dextran are prepared by mixing varying volume fractions of $11c^*$ DNA and dextran solutions, which we quantify by the volume fraction of the DNA solution $\phi$. We investigate samples with $\phi = 0$ (0% DNA, 100% dextran) $\phi = 0.25$ (25% DNA, 75% dextran), $\phi = 0.5$ (50% DNA, 50% dextran), $\phi = 0.75$ (75% DNA, 25% dextran), and $\phi = 1$ (100% DNA, 0% dextran). For each $\phi$, we prepare samples with ring DNA and linear DNA.

**2.4 Rheometry:** We use a Discovery Hybrid Rheometer 3 (DHR3, TA Instruments) with a 1º steel cone geometry to perform bulk rheology measurements. We use a glass slide as the bottom plate to enable imaging of fluorescent-labeled DNA in the samples. To prevent evaporation during the experimental cycle we apply mineral oil around the geometry and the sample. To measure linear viscoelastic moduli, $G'(\omega)$ and $G'(\omega)$, we perform two identical frequency sweeps from $\omega = 0.001$ to 100 rad/s at 5% strain (well within the linear regime as determined by amplitude sweeps). Each frequency sweep lasts ~6.5 hours with individual frequency measurements spaced 30 mins apart. All data shown is above the measurable torque minimum of 0.5 nN·m for the DHR3 rheometer.

**2.5 Fluorescence Microscopy:** The DHR3 is outfitted with a Modular Microscope Accessory (TA Instruments) with a 40× 0.6 NA objective (Nikon), blue-light LED source, 490/525 nm excitation/emission filters, and a Hamamatsu ORCA-Flash 2.8 CMOS camera to enable imaging of MPF488-labeled DNA in the blends. Immediately before and after each bulk rheology measurement, we collect three 512×512 pixel videos of 2000 frames at 1 fps. Sample frames from before and after videos are shown in Fig S2.

**2.6 Differential Dynamic Microscopy (DDM):** For DDM analysis (Fig 3), we split the videos into 256×256 pixel regions of interest (ROIs) which we analyze separately. We use custom-written scripts (Python) to perform DDM. For standard DDM analysis, one takes two-dimensional Fourier transforms of differences between images separated by a range of lag times $\Delta t$ in order to quantify how the degree of correlation decays with lag time as a function of the wave vector $q$. Because this standard correlation function is sensitive to global drift of the sample, we use a slightly modified correlation function referred to as the far-field DDM (FF-DDM) function. Previous work has shown that by using FF-DDM, the DDM correlation function, $D(q, \Delta t)$, is less sensitive to drift [73, 74]. As with standard DDM analysis, we fit the far-field DDM matrix to $D(q, \Delta t) = A(q)[1 - f(q, \Delta t)] + B(q)$, where $f(q, \Delta t)$ is the intermediate scattering function (ISF), $A(q)$ is the amplitude, and $B(q)$ is the background. To determine the type of motion and the corresponding rate, we model the ISF as a stretched exponential: $f(q, \Delta t) = e^{-(\Delta t/\tau(q))^\gamma}$ where $\tau(q)$ is the decay time and $\gamma$ the stretching exponent. The use of a stretched exponential as opposed to a simple exponential has been shown to better fit dynamics in confined or entangled systems [75, 76]. Scaling of $\tau(q) \sim q^{-2}$ is indicative of normal Brownian diffusion (i.e., $MSD \sim t$), whereas a decay time that depends less strongly on $q$ has been associated with more arrested or confined motion [77, 78]. Error bars shown in Fig 3 are standard error across all ROIs and videos, representing sample-to-sample variation. Example ISFs and fits are shown in Fig S4 and sample-to-sample variation in the ISFs is shown in Fig S5. Estimates of diffusion coefficients $D$ and their errors are based on the fits to $\tau(q) = 1/Dq^2$ over the range of $q$ from 0.39 μm$^{-1}$ to 2.44 μm$^{-1}$. For some samples, particularly for $\phi = 0.75$ composites, we observe a $q$-dependence that indicates nearly arrested rather than diffusive dynamics [78]. While we still compute an effective diffusion coefficient for these samples, because the dynamics differ from normal diffusive transport, we also fit each curve to $\tau(q) = 1/Kq^\alpha$ where $\alpha$ is a free parameter that is ~2 for normal diffusion and approaches zero for halted or arrested dynamics. $K$ is the generalized transport coefficient that is equivalent to $D$ when $\alpha = 2$.

**2.7 Single-molecule tracking**: For single-molecule tracking experiments, we image the MFP488-labeled DNA using an Olympus IX73 inverted fluorescence microscope with a 60× 1.2 NA oil immersion objective (Olympus). We collect five 512×512 pixel videos of 2000 frames at 10 fps for each sample. We use custom particle-tracking scripts (Python) to track the center-of-mass of individual DNA molecules and measure their frame-to-frame $x$ and $y$ displacements ($\Delta x, \Delta y$) from which we compute the ensemble averaged mean-squared displacements ($<\Delta x^2>, <\Delta y^2>$). We fit the average of $<\Delta x^2>$ and $<\Delta y^2>$ (i.e., $MSD$) versus lag time $\Delta t$ to a power-law function $MSD \sim \Delta t^\beta$ where $\beta$ is the subdiffusive scaling exponent. For a system exhibiting normal diffusion, $\beta=1$, while $\beta < 1$ indicates anomalous subdiffusion. Error bars listed in Fig 4 are based on fits of the data to $MSD \sim \Delta t^\beta$.

## 3. Results and Discussion

### 3.1 System Details

In all of our experiments described below, we fix the total polymer concentration to $11c^*$ to ensure that the molecules are highly overlapping, and we vary the volume fractions of the $11c^*$ solutions of the different polymers (ring DNA, linear DNA and dextran). By fixing the degree to which molecules overlap, we can unambiguously determine the effect of polymer topology (i.e., ring versus linear DNA), as well as steric and entropic interactions between distinct polymers (i.e., DNA and dextran), on the rheological properties and macromolecular dynamics. Specifically, we examine DNA-dextran blends with either purely linear DNA or 90%-10% ring-linear DNA (which we refer to as 'ring' throughout the paper for simplicity) at DNA volume fractions of $\phi = 0$ (pure dextran solution), 0.25, 0.5, 0.75 and 1 (pure DNA solution).

### 3.2 Bulk linear rheology of DNA-dextran composites

We first examine the bulk linear viscoelastic properties of DNA-dextran composites. Fig 2A compares the elastic modulus $G'(\omega)$ and viscous modulus $G''(\omega)$ for composites with linear DNA (left plot) and ring DNA (right plot) DNA. Pure dextran solutions (not shown) exhibit Newtonian properties with $G'' \sim \omega^1$ scaling over the entire frequency range. Linear DNA solutions without dextran ($\phi =1$) show a transition from the terminal flow regime at low frequencies, with approximate scaling of $G'' \sim \omega^1$ and $G' \sim \omega^2$, to a rubbery plateau at high frequencies with a crossover frequency, at which $G'$ exceeds $G''$, of $\omega_D \approx 0.25$ rad/s. This crossover frequency is a measure of the disengagement time, $\tau_D \approx 2\pi/\omega_D \approx 25$ s, which is the relaxation time associated with an entangled polymer reptating out of its confining tube. Different from linear DNA, ring DNA solutions ($\phi =1$) exhibit an elastic rubbery plateau over the entire frequency range, indicating very slow relaxation mechanisms at play ($\tau > 100$ mins), as previously reported and predicted for nominally pure ring melts and ring-linear blends with low fractions of linear chains [12, 17, 21-24, 29]. This slow relaxation has been predicted to arise from ring-ring and ring-linear threading which leads to kinetically arrested states [29, 30, 79].

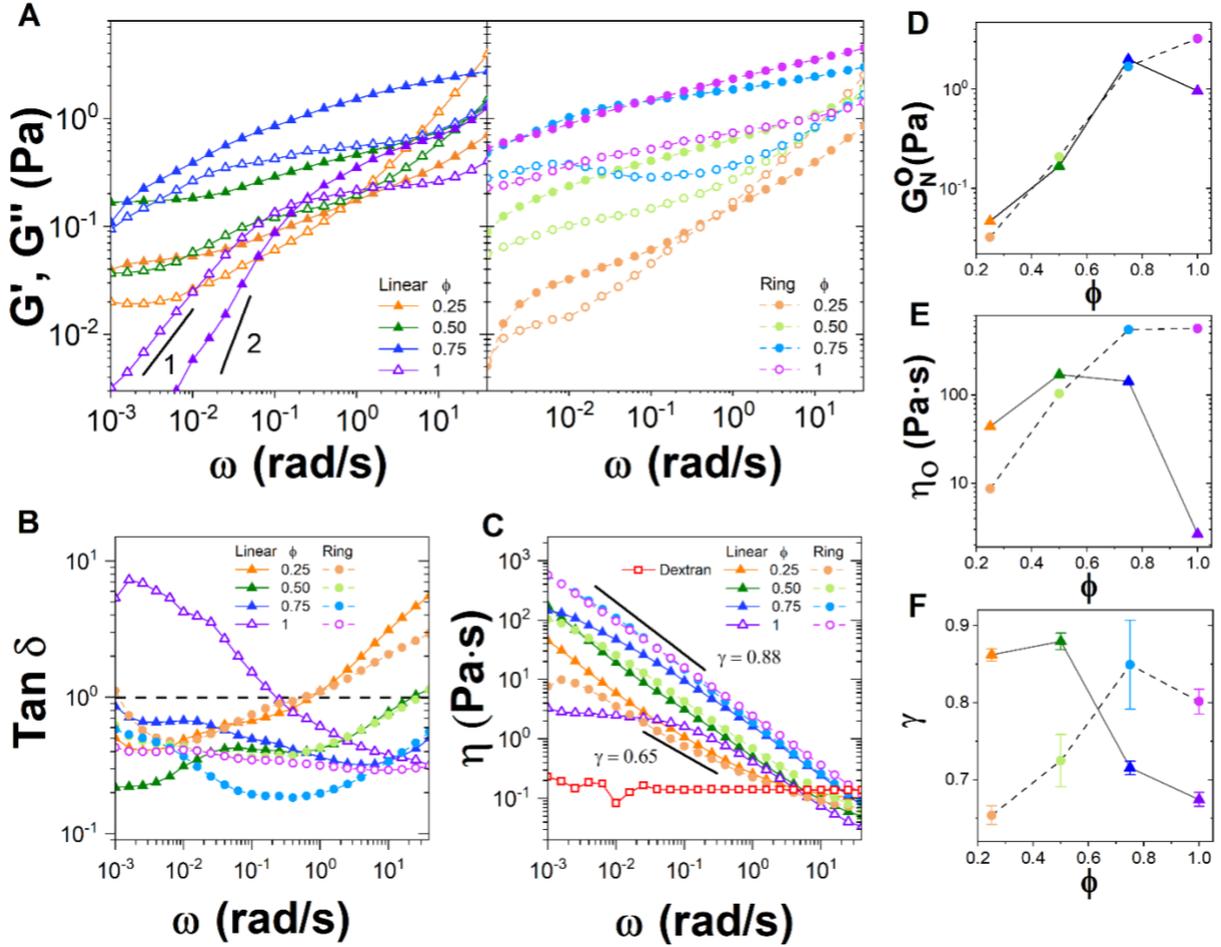

**Figure 2. Bulk linear rheology of DNA-dextran composites exhibits complex dependence on the topology and volume fraction of DNA.** (**A**) Elastic modulus $G'$ (closed symbols) and viscous modulus $G''$ (open symbols) as a function of frequency $\omega$ for DNA-dextran composites with linear DNA (left panel, triangles, dark shades) and ring DNA (right panel, circles, light shades) at volume fractions of $\phi = 0.25$ (orange), 0.5 (green) and 0.75 (cyan) and 1 (purple). Terminal regime scalings $G'(\omega) \sim \omega^{-2}$ and $G''(\omega) \sim \omega^{-1}$ are indicated by the scale bars in the left panel. (**B**) Loss tangent $\tan\delta = G''(\omega)/G'(\omega)$ for data shown in (A) with open symbols denoting pure DNA solutions ($\phi = 1$). Dashed line denotes $\tan\delta = 1$ which corresponds to low and high crossover frequencies, $\omega_D = 2\pi/\tau_D$ and $\omega_e = 2\pi/\tau_e$, respectively. (**C**) Complex viscosity $\eta(\omega) = [G''(\omega)^2 + G'(\omega)^2]^{1/2}/\omega$ of composites shown in (B) and a pure dextran solution ($\phi = 0$, red open squares). All composites exhibit shear thinning $\eta(\omega) \sim \omega^{-\gamma}$ with exponents $\gamma$ that vary from 0.65 to 0.88. (**D**) Plateau modulus $G_N^0$ versus $\phi$ for composites with ring (circles, dashed connecting line) and linear (triangles, solid connecting line) DNA. $G_N^0$ is computed by evaluating $G'$ at the frequency in which $\tan\delta$ is a minimum. (**E**) Zero-shear viscosity $\eta_0 = \eta(\omega \to 0)$ versus $\phi$ for composites with ring (circles, dashed connecting line) and linear (triangles, solid connecting line) DNA determined from $\eta(\omega)$ curves shown in (C). (**F**) Shear thinning exponent $\gamma$ versus $\phi$ determined from power-law fits to the data shown in (C). Error bars are determined from fits to the data.

In contrast to previous studies on ring and linear polymer melts [19, 21], the magnitude of the plateau modulus $G_N^0$ for ring DNA is several times larger than that for linear DNA. To shed light on this discrepancy, we quantify $G_N^0$ for all solutions by evaluating $G'$ at the frequency at which the corresponding loss tangent $\tan\delta = G''(\omega)/G'(\omega)$ (Fig 2B) is a minimum [80]. We find $G_N^0 \simeq$ 3.24 Pa and ~0.96 Pa for ring and linear DNA solutions, respectively, such that $G_{N,R}^0/G_{N,L}^0 \simeq 3.4$ (Fig 2D). From $G_N^0$ we can estimate the polymer length between entanglements $l_e$ via $l_e = 4cRT/5G_N^0 M_{BP}$ where $c$ is the mass concentration and $M_{BP} \approx 650$ g/mol is the molecular weight of a DNA basepair (~1/3 nm) [8, 27]. For linear DNA, we find $l_e \simeq 833$ nm and the number of entanglements per chain $N_e \approx 10$. As described in Methods, to maintain concentration at $11c^*$, the mass concentration of the nominal ring sample is ~3-fold higher than for the linear sample ($11c_{0.9R}^*/11c_L^* \simeq 2.42/0.78 \simeq 3.1$) due to the smaller $R_G$ of rings. As $l_e \sim c/G_N^0$, the ~3-fold larger $G_N^0$ and $c$ for rings is consistent with the fact that we are fixing the degree of polymer overlap and thus $l_e$. Thus, while a ~3.4-fold higher $G_N^0$ value is, at first glance, at odds with previous studies [19, 21, 31], the increased mass concentration of our rings compared to linear DNA accounts for this increase.

Upon mixing DNA with dextran (thereby decreasing $\phi$), we find that $G_N^0$ decreases for rings, while for linear DNA-dextran composites, $G_N^0$ displays a non-monotonic dependence on $\phi$ with a maximum at $\phi$=0.75. For both topologies, $\phi \leq 0.5$ blends display a high frequency crossover $\omega_e$ in which $G'' > G'$, which is a measure of the entanglement time $\tau_e \approx 2\pi/\omega_e$ or the time it takes for an entangled polymer to 'feel' its tube confinement [8]. This crossover frequency increases from ~0.63 at $\phi$ =0.25 to ~25 rad/s at $\phi$ =0.5, independent of DNA topology, and is non-existent for $\phi$ = 0.75 and 1 (implying $\omega_e$>100 rad/s). These values correspond to $\tau_e \approx 10$ s, 0.25 s and <0.06 s, for $\phi$ =0.25, 0.5 and >0.5.

The independence of $\tau_e$ on topology can be understood as arising from the matching of $l_e$. Namely, $\tau_e \sim a^4/D_0 R_{G,0}^2$ where $a \approx (l_K l_e)^{1/2}$ is the entanglement tube diameter, $l_K$ is the Kuhn length (~100 nm for dsDNA), and $D_0$ and $R_{G,0}$ are the diffusion coefficient and radius of gyration in the dilute limit [8, 27, 81]. Insofar as $D_{0,R}/D_{0,L} \approx 1.32$, $R_{G,0,L}/R_{G,0,R} \approx 1.58$ [68], and $l_{e,L} \approx l_{e,R}$, we find $\tau_{e,R} \approx 1.8\tau_{e,L}$; such that, within the factor of 2 frequency resolution of our measurements, we can estimate $\tau_{e,R} \approx \tau_{e,L}$.

Further, if we only consider the DNA in the composites, and use the relations $\tau_e \sim l_e^2$, $l_e \sim c/G_N^0$ and $G_N^0 \sim c^2$ [8], it follows that $l_e$ decreases as $\phi$ increases (i.e., $l_e \sim c^{-1} \sim \phi^{-1}$) so $\tau_e$ is likewise expected to decrease, as we find in composites. The decrease in $\tau_e$ between $\phi = 0.25$ and 0.50 is stronger than the predicted scaling, suggesting that dextran may serve to increase the entanglement density and overlap of DNA via depletion interactions. To incorporate the effect of the dextran, we use our estimated $G_N^0$ values (Fig 2D), along with $l_e \sim \phi/G_N^0$ and $l_{e,DNA} \approx 833$ nm, to estimate $l_e$ for the different composites. This analysis (Fig S3) shows that, for linear DNA, the $\phi = 0.75$ composites have the smallest predicted $l_e$ (~300 nm), whereas for rings, $\phi = 1$ solutions have the lowest (~833 nm). Taken together, these results suggest that the peak in $G_N^0$ at $\phi = 0.75$ for linear

DNA may be due to depletion-driven DNA self-association and increased overlap in the presence of dextran, which plays a lesser role for ring DNA.

Further examination of the loss tangent (Fig 2B) shows that, for linear DNA composites, the rubbery plateau, flanked by the low and high frequencies at which $\tan\delta = 1$ (i.e., $\omega_D$ and $\omega_e$) shifts to longer timescales (lower $\omega$) and extends over a wider $\omega$ range when mixed with dextran. We also see that $\tan\delta$ generally increases with $\omega$ for $\phi \leq 0.5$ composites whereas the opposite is true for $\phi > 0.5$. This qualitatively different $\omega$ dependence is reminiscent of the difference between rheological properties of extended semiflexible polymers, such as actin filaments, which exhibit solid-like behavior at low $\omega$ and viscous-dominated flow at high $\omega$ [82], compared to entangled flexible random coil polymers (e.g., DNA) which display low-$\omega$ flow and high-$\omega$ rubber-like elasticity. This effect may indicate that dextran polymers are acting as depletants to extend and bundle linear DNA molecules, making them more akin to extended semiflexible polymers [54, 55, 83]. As the DNA concentration increases and the dextran concentration decreases the depletion-driven bundling becomes weaker while DNA-DNA entanglements are more abundant so the frequency-dependence is more akin to that of flexible polymers but the modulus is higher. Thus, the maximum in $G^0_{N,L}$ at $\phi = 0.75$, suggests an optimal combination of DNA overlap and depletion-driven stretching and self-association. Below $\phi = 0.75$, the DNA in the composite has fewer entanglements (Fig S3), so reinforcement from bundling comes at the cost of losing connectivity, thereby reducing $G^0_{N,L}$.

For ring composites, the rubbery plateau is truncated at short times (high $\omega$) in $\phi = 0.25$ and 0.5 composites compared to $\phi = 0.75$ and 1 cases in which the plateau regime fully spans over four decades. Further, compared to linear DNA composites, there is limited $\phi$ dependence at low $\omega$, suggesting that bundling and elongation play minimal roles in ring composites. We suggest instead that the variation in threading propensity of rings dominates the rheology for ring composites. Previously, we found that high dextran concentrations (>$10c^*$) led to compaction of ring DNA, rather than stretching,[83] which would result in fewer entanglements and more viscous-dominated rheology compared to a ring DNA solution at the same degree of overlap, as shown in Fig 2A. Further, due to >10x smaller coil size of dextran compared to DNA rings, the ability for dextran to kinetically pin rings via threading is likely minimal. Indeed, ring melts that contain a small fraction of short polymer chains have been shown to have reduced viscosity, owing to fewer entanglements, while at the same time the rings swell due to linear chain-end effects [40].

Such swelling, which may occur for higher DNA volume fractions (versus compaction observed for higher dextran fractions), likely facilitates ring-ring and ring-linear DNA threading in composites that suppress relaxation. Swelling will also limit the fraction of double-folded and amoeba-like ring conformations that can undergo modified reptation but are not threaded [25, 84]. Thus, in the $\phi = 0.75$ ring composite the relaxation is likely entirely due to constraint release of threaded DNA, while the $\phi = 1$ DNA solution has both reptation and threading contributions. This effect can be seen in Fig 2B in which $\tan\delta$ for the $\phi = 0.75$ ring composite is smaller than that for $\phi = 1$ over most of the frequency range ($\omega \approx 0.006 - 6$ rad/s), indicating the timescale over which threading, which occurs more frequently at $\phi = 0.75$ than $\phi = 1$, dominates ($t_{th} \approx 1$ to $\sim 10^{-3}$ s). Above and below this timescale, unthreaded and unswollen rings in $\phi = 1$ solutions may still 'feel' the constraints of the entanglement tube (no measurable $\tau_e$) and be confined to reptate within it (no measurable $\tau_D$).

Finally, we evaluate the complex viscosity $\eta(\omega)$ (Fig 2C), and, assuming that the Cox-Merz rule is valid for our solutions [16, 18, 33, 85-87], we estimate the zero-shear viscosity $\eta_0 = \eta(\omega \to 0)$ (Fig 2E) and corresponding shear-thinning exponent $\gamma$ (Fig 2F) by fitting each curve to a power-law $\eta(\omega) \sim \omega^{-\gamma}$. Because only the linear DNA solution and $\phi=0.25$ ring composite exhibit low-frequency plateaus, our $\eta_0$ values for the other cases should be taken as lower limits. As shown in Fig 2D we find that $\eta_0$ for linear DNA composites exhibit a non-monotonic dependence on $\phi$, corroborating our interpretation of depletion-driven stretching and bundling of linear DNA in composites.

Conversely, for rings, $\eta_0$ increases monotonically with $\phi$, in agreement with previous studies showing that dilute rings in a melt of short linear chains (akin to low $\phi$ composites) have lower viscosity than for pure rings [40]. Further $\eta_0$ is maximum at $\phi=1$ rather than $\phi=0.75$ because $\eta_0$ is evaluated in the long-time limit, such that $t > t_{th}$, so threading events are no longer constraining the rings but tube confinement is still present, as indicated by Fig 2B. The latter is more prevalent for $\phi=1$ ring solutions compared to $\phi=0.75$ composites.

We next examine the shear thinning exponents (Fig 2F), which provide insight into the ability of entangled polymers to align with flow [18, 88, 89]. For highly entangled linear polymer melts, $\eta(\omega) \sim \omega^{-1}$, while shear-thinning exponents of $\gamma \approx 0.6 – 0.9$ have been reported for solutions of entangled linear DNA [16, 20]. Previous studies on entangled rings report reduced thinning exponents of $\gamma \approx 0.4 – 0.6$ [16, 20], owing to the reduced ability of rings to stretch along the shear direction. Our previous studies on ring-linear DNA blends showed a maximum in $\gamma$ with comparable concentrations of rings and linear chains, in which threading was most pervasive, suggesting that threading facilitates flow alignment of rings [16]. Moreover, if we assume rings are compacted by dextran at low $\phi$ while linear chains are stretched and bundled, we would expect minimal flow alignment and thinning for rings and maximal alignment for linear chains at $\phi=0.25$ and 0.5, which is indeed what we see. For $\phi > 0.5$, $\gamma$ drops significantly for linear DNA whereas it increases a comparable amount for ring DNA. Thus, for rings, threading, mediated by chain swelling, appears to promote thinning whereas for linear chains it is the depletion-driven elongation and increased overlap.

### 3.2 Differential Dynamic Microscopy to determine macromolecular transport properties

To shed light on the macromolecular dynamics that give rise to the bulk rheological properties discussed above, we use an optical microscopy attachment to our rheometer to collect time-series of images of diffusing linear and ring DNA in the composites (Fig S2), and perform DDM analysis on the time-series to determine their transport properties (Fig 3) [90, 91]. Importantly, these data are collected in the exact samples and geometry as bulk rheology data immediately before and after rheology measurements (see Methods).

Some key questions we seek to answer with this analysis are: (1) if ring DNA solutions and composites exhibit glassy dynamics, (2) if mixing of linear DNA with dextran slows its diffusion and/or alters its transport properties, (3) the extent to which the relaxation and diffusive mechanisms that rings and linear chains exhibit in the different conditions are similar or different, and (4) if the topology and $\phi$ dependence of bulk rheological properties manifest at the molecular level or if there are scale-dependent phenomena that arise, as seen in other entangled polymer solutions and composites [33, 65, 92-94].

Fig 3A shows the characteristic decorrelation times $\tau$ of diffusing DNA polymers versus wavenumber $q$, which we determine by fitting the radially-averaged image structure function $D(q, \Delta t)$ to a stretched exponential, as described in Methods. In general, higher $\tau$ values for a given $q$ indicate slower motion. By fitting $\tau(q)$ to the power-law function $\tau = 1/(Kq^\alpha)$, one can determine the type of motion. Normal Brownian diffusion is described by $\alpha = 2$ with $K$ equating to the corresponding diffusion coefficient $D$. We can therefore estimate diffusion coefficients for our data by fitting each curve to $\alpha = 2$ scaling (Fig 2C). Conversely, restricted or halted motion, such as in glassy systems, often results in much weaker $q$ dependence ($\alpha \to 0$) as we see in some of the data (Fig 3D) [77, 78]. If $\alpha$ deviates from 2 then $K$ is a generalized transport coefficient with dimensions that depend on $\alpha$ (Fig 3E).

For both topologies, transport is fastest (smallest $\tau$) and scaling is closest to diffusive ($\alpha = 2$) in dextran solutions ($\phi$=0), and is slowest and exhibits the weakest dependence in $\phi$=0.75 composites. To better visualize the deviation from normal diffusion and the non-monotonic $\phi$ dependence shown in Fig 3A, we plot $1/\tau q^2$ (Fig 3B), which, for diffusive transport, is a horizontal line with a $y$-intercept equal to $D$ (Fig 3C). Halted transport or kinetic arrest manifests as a negative slope approaching $\alpha$-2$\approx$-2, as seen for $\phi$=0.75 ring composites. Such kinetically arrested states, predicted by simulations of ring polymer melts, indicate threading events.

It may be possible that ring DNA is threaded by dextran polymers. However, the much smaller size of dextran polymers compared to DNA rings, would limit their ability to halt ring diffusion via constraint release. Indeed, as previously observed in ring-linear melts, short linear polymers (akin to the dextran used here) cause tracer rings to swell and the blends exhibit reduced viscosity compared to ring melts, in contrast to when the linear chains are comparable in size to the rings [40]. Dextran-mediated swelling at $\phi = 0.75$ could facilitate DNA-DNA threading events responsible for glassy behavior, by reducing the fraction of double-folded and collapsed rings [25].

These results corroborate our bulk rheology data that suggests that threading of rings is most pervasive at $\phi$=0.75. Conversely, $\phi$=1 and 0.5 ring composites exhibit $q$-dependence that is in between diffusive and arrested ($\alpha \approx 1.4$) (Fig 3D), indicative of contributions from populations of both diffusive entangled rings and halted threaded rings–as our rheology results suggest.

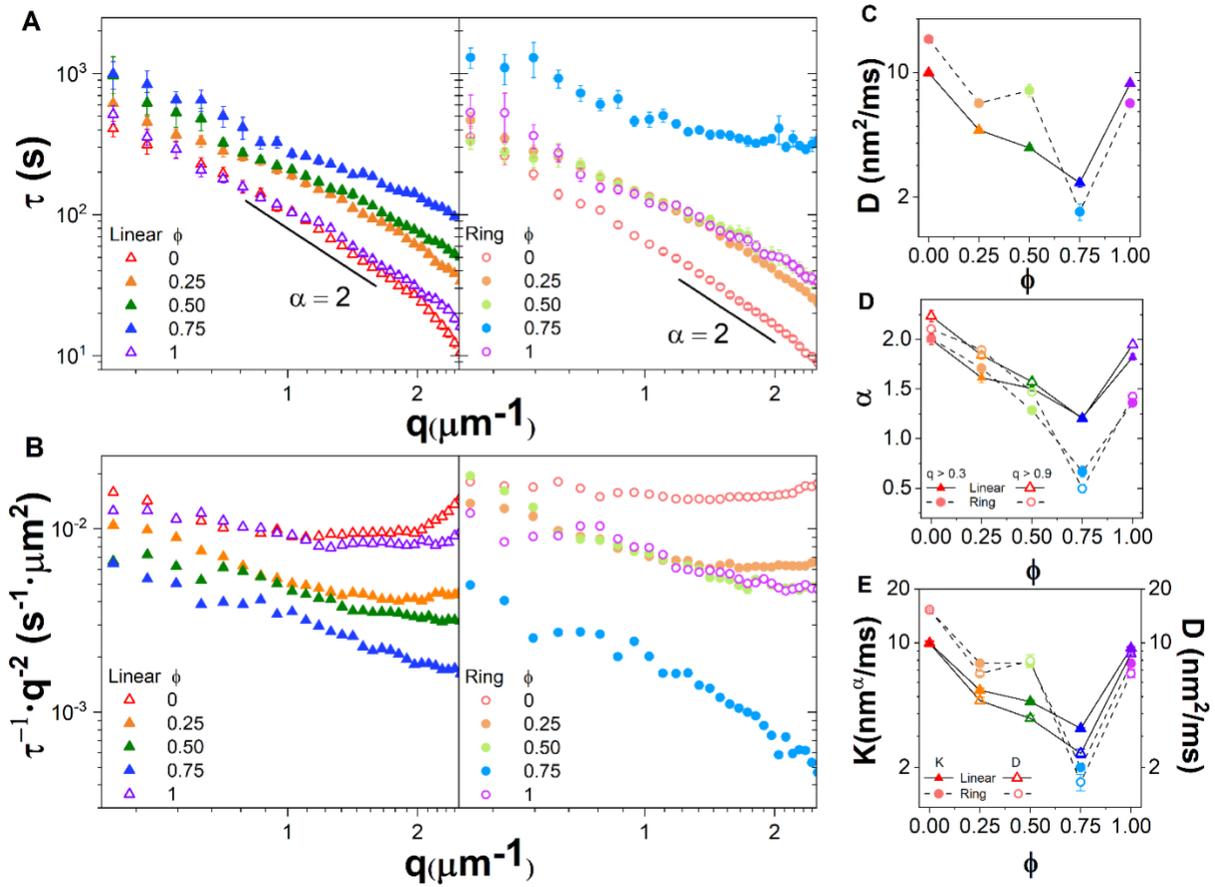

**Figure 3. Differential dynamic microscopy of DNA-dextran composites show arrested dynamics and non-monotonic dependence on the volume fraction of DNA.** (**A**) Characteristic decay time $\tau$ versus wave vector $q$ for DNA-dextran composites with linear DNA (left panel, triangles, dark shades) and ring DNA (right panel, circles, light shades) at volume fractions of $\phi = 0$ (red), $\phi = 0.25$ (orange), 0.5 (green) and 0.75 (cyan) and 1 (purple). Scaling bars indicate diffusive dynamics, i.e., $\tau(q) \sim q^{-2}$. (**B**) Data shown in (A) plotted as $\tau^{-1}q^{-2}$, which equates to a $q$-independent diffusion coefficient $D$ for systems that exhibit normal diffusion. Increased negative slopes indicate more arrested or halted motion. (**C**) Diffusion coefficient versus $\phi$ for composites with ring (circles, dashed connecting line) and linear (triangles, solid connecting line) DNA determined by fitting the data in A to $\tau(q) = 1/(Dq^2)$. (**D**) Transport scaling exponent $\alpha$ versus $\phi$ for composites with ring (circles, dashed connecting line) and linear (triangles, solid connecting line) DNA determined from fitting the data in A to $\tau(q) = 1/(Kq^\alpha)$. Fits are performed over the entire $q$ range shown (filled symbols) and for $q > 0.9$ (open symbols). $q=2$ signifies normal diffusion and $q \rightarrow 0$ indicates kinetic arrest. (**E**) Generalized transport coefficient $K$ (left axis, filled symbols) compared to the corresponding diffusion coefficient $D$ (right axis, open symbols) versus $\phi$ for composites with ring (circles, dashed connecting line) and linear (triangles, solid connecting line) DNA determined from fitting the data in (A) to $\tau(q) = 1/(Kq^\alpha)$ or $\tau(q) = 1/(Dq^2)$. Error bars (many of which are too small to see) are determined from fits to the data.

Linear DNA composites exhibit similar trends, with minimum $\alpha$ and $D$ values at $\phi$=0.75, followed by 0.5, but the variation in $\alpha$ with $\phi$ is much weaker (Fig 3C,D). This weaker $\phi$-dependence for linear chains is more evident when we evaluate the generalized transport coefficients $K$ for the different composites: $K$ exhibits a weaker $\phi$-dependence for linear chains than the corresponding diffusion coefficient (Fig 3E). Further, the pure linear DNA solution exhibits diffusive scaling ($\alpha \approx 2$), in contrast to the rings ($\alpha \approx 1.4$). These topology-dependent effects highlight the distinct relaxation mechanisms at play in the different composites. Namely, threading dominates ring composites whereas depletion-driven DNA stretching and self-association mediate the restricted diffusion in linear DNA composites.

These different transport modes are expected to have different degrees of microscale heterogeneity [95, 96], which we evaluate by examining the spread in the intermediate scattering functions (ISF) generated from DDM performed on different ROIs in the time-series (Fig S5). In general, the spread in the ISF for rings is larger than that for linear composites, in line with simulations that show that topological glasses exhibit heterogeneous transport [29, 30]. Moreover, considering the ring composites, the spread in the ISFs for $\phi$=0.5 and 1 are significantly broader than that for $\phi$=0.75, indicating that the $\phi$=0.75 composite is dominated by a single transport mode (i.e., threading) whereas $\phi$=0.5 and $\phi$=1 composites undergo a combination of modes (i.e., threading and modified reptation) which occur over varying timescales.

Finally, we note that the scaling of $\tau(q)$ is weaker at smaller $q$ values (i.e., longer lengthscales $\lambda$), as depicted by comparing $\alpha$ values computed over the entire $q$ range ($q = 2\pi/\lambda \approx 0.3 - 2.5$ $\mu m^{-1}, \lambda \approx 2.5 - 20$ $\mu$m) to those excluding $q_c \lesssim 0.9$ $\mu m^{-1}$ ($\lambda_c \gtrsim 7$ $\mu$m) (Fig 3D). We can understand this effect and the corresponding lengthscale $\lambda_c$ as arising from entanglements and threadings that hinder polymer motion beyond a few entanglement lengths $l_e$ [95, 97], such that restricted transport ($\alpha \to 0$) will be more apparent for $\lambda > O(l_e)$. Indeed, Fig S3 shows that the range of $l_e$ values for the different composites span from ~0.3 $\mu$m (for linear $\phi$=0.75) to 20 $\mu$m (for ring $\phi$=0.25), on the order of $\lambda_c$, corroborating this interpretation.

### 3.3 Single-molecule tracking to corroborate DDM and expand the spatiotemporal scales of transport measurements

To further understand the non-monotonic dependence of transport and rheology on $\phi$, and the $\phi$-dependent differences between ring and linear DNA, we perform single-molecule tracking measurements of DNA in $\phi = 0.75$ and $\phi = 1$ composites. We focus on $\phi = 0.75$ as it deviates most strongly from normal diffusive scaling for both topologies. Fig 5 shows the center-of-mass mean-squared displacements ($MSD$) versus lag time $\Delta t$ for ring and linear DNA at $\phi = 0.75$ and 1. All cases exhibit anomalous subdiffusion, i.e., $MSD \sim \Delta t^\beta$ where $\beta <1$. For both $\phi$ values, rings display greater deviation from normal diffusion compared to linear DNA, with an average value of $\beta \simeq 0.34$ compared to ~0.67 for linear chains. These values are comparable to, but slightly lower than, those previously reported for tracer ring and linear DNA molecules in networks of cytoskeleton filaments [93, 95, 97]. In those experiments the lower $\beta$ value for rings was attributed to threading by cytoskeleton filaments, further supporting our interpretation of our results shown in Figs 2 and 3. Notably, previous tracking experiments of tracer ring and linear DNA in concentrated dextran solutions reported normal diffusion [56, 83], similar to our DDM measurements (Fig 3), supporting our conclusion that threading by dextran does not contribute to the phenomena we report here, and that subdiffusion of linear DNA in composites arises from DNA-DNA bundling which requires a critical fraction of DNA. Subdiffusive transport, with similar exponents to our linear DNA composites, has also been reported for entangled solutions of stiffer and more extended actin filaments [98, 99], further indicating bundling and stretching as the mechanism underlying the hindered diffusion and Increased elasticity of linear DNA composites compared to monodisperse DNA solutions.

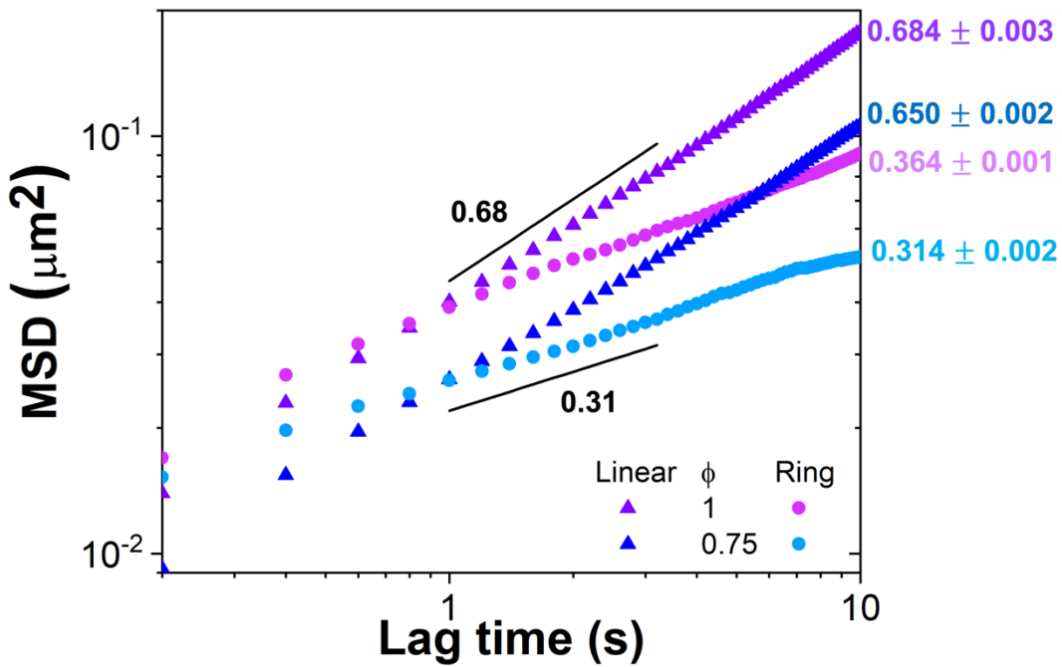

**Figure 4. Ring and linear DNA exhibit topology-dependent subdiffusion at high DNA volume fractions.** *MSD* versus lag time $\Delta t$ for linear (triangles) and ring (circles) DNA diffusing in composites with DNA volume fractions of $\phi = 1$ (dark and light purple) and $\phi = 0.75$ (dark and light blue). Black lines represent power-law scaling with exponents listed. The exponent $\beta$ from fitting each curve to $MSD \sim \Delta t^{\beta}$ is listed to the right of each curve. In all cases, DNA exhibits subdiffusion ($\beta < 1$) with rings displaying greater deviation from normal diffusion ($\beta = 1$) than linear DNA. For both topologies *MSD*s are ~2x lower in $\phi = 0.75$ composites compared to $\phi = 1$.

Further, $\beta$ is slightly lower (outside of measured error) for $\phi = 0.75$ compared to $\phi = 1$ for both topologies, and the value of the *MSD* at a given lag time is ~2-fold lower, in line with our DDM and bulk rheology data. The strongly anomalous diffusion and concomitantly low $MSD$ for rings at $\phi = 0.75$ corroborate our interpretation of glassy dynamics due to threading.

Finally, we note that the timescale probed by single-molecule tracking is lower than for DDM (Fig 1B), corresponding to $\omega \simeq 0.6$ - 30 rad/s. In this range, the linear DNA solution ($\phi$=1) exhibits more elastic-like behavior compared to lower frequencies, probed by DDM, in which it exhibits terminal flow behavior. As such, DDM measurements show $\phi = 1$ linear DNA obeying normal Brownian diffusion ($\alpha \approx 2$) while single-molecule tracking experiments show subdiffusive transport.

### 3.4 Coupling bulk rheological properties to macromolecular dynamics

To quantitatively compare and couple bulk and molecular-level dynamics we compare the diffusion coefficient determined from DDM (Fig 3D) to the zero-shear viscosity measured via bulk rheology (Fig 2E). In Newtonian fluids (as well as some non-Newtonian fluids) these quantities are inversely related via the Stokes-Einstein fluctuation-dissipation relation, such that $D$ vs $\eta_0^{-1}$ should exhibit linear scaling, provided the size and conformation of the polymers are not changing. As shown in Fig 5A, we find that, in general, higher diffusion coefficients are coupled to lower viscosities, as we may expect. However, examination of $D$ vs $\eta_0^{-1}$ reveals regions of the phase space in which the metrics of mobility at the molecular-level ($D$) and bulk-level ($\eta_0^{-1}$) do not scale together (Fig 6C). For example, ring composites with $\phi$=0.25, 0.5 and 1 have nearly identical diffusion coefficients yet the corresponding $\eta_0^{-1}$ values vary by two orders of magnitude. Conversely, it is clear that the $\phi$=0.75 ring composite is the least mobile and the $\phi$=1 linear DNA is the most mobile across all scales.

We also compare the bulk shear thinning exponent $\gamma$ with the deviation of the DDM transport coefficient from diffusive dynamics, which we define as the quantity 2-$\alpha$ (Fig 6B). Both metrics quantify constraint, with $\gamma$ ranging from 0 for Newtonian fluids to 1 for highly entangled or threaded systems and 2-$\alpha$ spanning from 0 for normal Brownian diffusion to 2 for arrested dynamics. As such, we expect the $\phi$ and topology dependence of $\gamma$ and 2-$\alpha$ to track with one another. This relationship loosely holds for many of the composites, with the $\phi$=0.75 ring composite appearing the most constrained and the $\phi$=1 linear composite exhibiting negligible constraints across all scales. However, we also find cases, such as $\phi$=0.25 and 0.5 linear DNA composites, that appear highly entangled and constrained at the bulk scale while displaying relatively little deviation from diffusive scaling at the molecular scale.

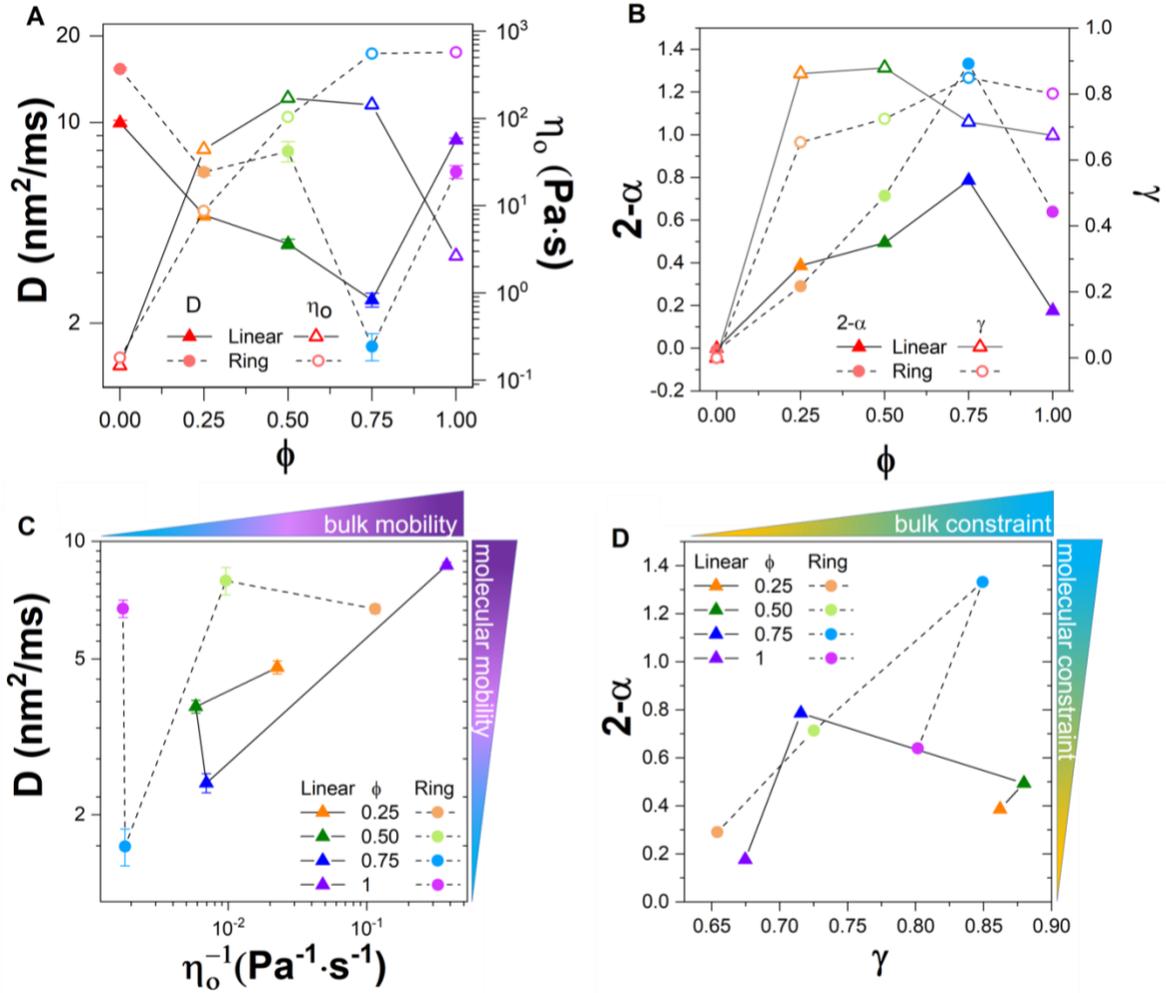

**Figure 5. Coupling of bulk and molecular-level metrics reveals emergent scale-dependent dynamics in DNA-dextran composites.** (**A**) Diffusion coefficient *D* determined from DDM (left axis, closed symbols) and zero-shear viscosity $\eta_0$ (right axis, open symbols) determined via bulk rheology for DNA-dextran composites with linear (triangles) or ring (circles) DNA at varying DNA volume fractions $\phi$. (**B**) Deviation from diffusive DDM scaling, quantified by 2-$\alpha$ (left axis, closed symbols), and shear-thinning exponent $\gamma$ (right axis, open symbols) determined via bulk rheology versus $\phi$ for DNA-dextran composites with linear (triangles) and ring (circles) DNA. (**C**) Data shown in (A) plotted as *D* versus $\eta_0^{-1}$ for DNA-dextran composites with ring (circles, dashed connecting line) and linear (triangles, solid connecting line) DNA. Both quantities are measures of mobility with larger values indicating higher mobility, as indicated by the gradients. (**D**) Data shown in (B) plotted as 2-$\alpha$ versus $\gamma$ for DNA-dextran composites with ring (circles, dashed connecting line) and linear (triangles, solid connecting line) DNA. Both quantities are indicators of steric constraints with larger values indicating more constrained dynamics, as indicated by the gradients.

Finally, the data in Fig 5 show that $\phi$=0.25 linear DNA composites have increased bulk $\eta_0$ and $\gamma$ values and reduced microscale *D* and $\alpha$ values compared to $\phi$=1 linear DNA solutions, while the opposite is true of all metrics for rings. We understand this effect as arising from the loss of DNA-DNA threadings which, when present, lead to kinetically arrested states that suppress bulk dissipation. Further, dextran solutions at sufficiently high volume fractions (comparable to $\phi \leq 0.5$ composites) have been shown to compact ring DNA which would serve to reduce entanglements with neighbors at low DNA concentration. Indeed, we show that $l_e$ for rings at $\phi=0.25$ is greater than the DNA contour length $L$ and greater than that computed assuming $l_e \sim \phi^{-1}$ scaling, indicating that compaction and reduced entanglements dictate ring dynamics at low $\phi$. Conversely, recent work demonstrates that rings are swollen by short linear polymers when the linear polymers are below the entanglement threshold (such as in our high $\phi$ composites) [40]. This swelling, like the stretching of linear DNA, promotes entanglements and threading, which likewise facilitates shear thinning (Fig 2F). As such, we argue that rings switching between compacted and swollen states, underlies the phenomenon that the diffusion coefficient $D$ is lower, and the zero-shear viscosity $\eta_0$ and shear thinning exponent $\gamma$ are higher, for linear versus ring DNA composites at $0 \leq \phi \leq 0.5$, whereas rings become slower and exhibit enhanced bulk viscosity and shear thinning compared their linear counterparts beyond $\phi=0.5$.

These emergent regions of the phase space, indicate that composites can exhibit scale-dependent dynamics tuned by the relative concentrations of DNA and dextran as well as the DNA topology. These data also demonstrate the need for coupled microscale and bulk measurement methods to shed light on the physics underlying the rheology and transport properties that complex systems manifest. A final example to demonstrate this point is the apparent discrepancy between $D$ measured via DDM and $\tau_D$ measured via bulk rheology for $\phi=1$ linear DNA. For entangled linear polymers, $\tau_D \approx L_0^2/D$ where $L_0 \approx (L/l_e)a$ is the primitive path length of the polymer [8]. For our linear DNA solution, we measure $D \approx 9$ nm²/ms via DDM, within a factor of two of the value measured previously with single-molecule tracking [69]. Using this value gives $\tau_{D,micro} \approx 920$ s, which is substantially larger than $\tau_{D,bulk} \approx 25$ s measured via bulk rheology. However, because our entanglement density of $Z \approx L/l_e \approx 10$ (determined from $G_N^0$, see Figs S3 and 2D) is well below O(10²), we cannot ignore contour length fluctuations (CLF), i.e., fluctuations in the length of the primitive path, which reduce the disengagement time [8]. Specifically, CLF contributions are predicted to reduce $\tau_D$ according to the relation $\tau_D^{(CLF)} \approx \tau_D \left(1 - \frac{X}{\sqrt{Z}}\right)^2$ where $X$ is a numerical factor that is of order unity and predicted to be greater than ~1.5 [8, 100]. Using $\tau_D^{(CLF)} \approx \tau_{D,bulk} \approx 25$ s and $\tau_D \approx \tau_{D,micro} \approx 920$ s, we compute $X \approx 2.6$, which is indeed of order unity and >1.5. Thus, not only do our bulk and molecular-level measurements align with one another, but, taken together, they demonstrate the significance of CLF contributions to entangled chain dynamics and shed light on the numerical factor that relates the two for entangled DNA.

**Conclusions**

In summary, we have coupled bulk rheology measurements with fluorescence imaging and DDM analysis to directly correlate the bulk viscoelastic properties of entangled DNA-dextran composites to the corresponding microscale polymer transport (Fig 1). Importantly, we performed both

measurements – at lengths scales that differ by ~4 orders of magnitude–on the exact same samples, under the same conditions, and within minutes of each other. In this way, we explicitly connect macroscopic rheological properties to the underlying macromolecular dynamics (Fig 5).

We show that DNA-dextran composites exhibit topology-dependent non-monotonic dependences of both bulk rheological properties and macromolecular transport properties on the fraction of DNA comprising the composites. We attribute this emergent behavior to the different ways in which dextran polymers alter the conformation and self-association of ring and linear DNA via depletion interactions. Rings may be either compacted (at $\phi \leq 0.5$) or swollen (at $\phi = 0.75$) by dextran. When swollen, threadings by neighboring DNA give rise to near kinetic arrest. For linear DNA, depletion-driven stretching and bundling of the DNA may either reduce connectivity (at low $\phi$) or give rise to overlapping extended bundles of DNA with properties akin to semiflexible polymers (at high $\phi$).

More generally, our results demonstrate that for polymer composites, the whole is not always equal to the sum of its parts, rather mixing polymers with distinct structures, sizes and topologies can give rise to emergent dynamics that can either span from microscopic to macroscopic scales or display scale-dependence depending on the composite composition. We further show that polymer end-closure plays an important role in interactions between the different species comprising the composites, with ring polymers conferring uniquely suppressed dissipation and relaxation.

## Supplementary Materials

We provide the following figures in Supplementary Materials: Fig S1. Gel electrophoresis of linear and ring DNA solutions; Fig S2. Sample images of labeled DNA from time-series recorded before and after bulk rheology measurements; Fig S3. Nominal entanglement lengths computed from the plateau modulus for DNA in DNA-dextran composites; Fig S4. Sample DDM intermediate scattering functions for DNA-dextran composites; Fig S5. Variations in the intermediate scattering function across different regions of interest.

## Acknowledgements

We acknowledge Maya Nugent for her help in preparing figures, and acknowledge funding from AFOSR awards (FA9550-17-1-0249, FA9550-21-1-0361) to RMRA and an NSF CBET award (CBET-1919429) to RJM (PI) and RMRA (co-PI).

## Conflicts of Interest

The authors have no conflicts to disclose